# THE SIX-CAVITY TEST - DEMONSTRATED ACCELERATION OF BEAM WITH MULTIPLE RF CAVITIES AND A SINGLE KLYSTRON*

J. Steimel#, J.P. Carneiro, B. Chase, E. Cullerton, B. Hanna, R. Madrak, R. Pasquinelli, L.R. Prost, L. Ristori, V. Scarpine, P. Varghese, R. Webber, D. Wildman, FNAL, Batavia, IL 60510, USA.

*Abstract*

The High Intensity Neutrino Source (HINS) 'Six-Cavity Test' has demonstrated the use of high power RF vector modulators to control multiple RF cavities driven by a single high power klystron to accelerate a non-relativistic beam. Installation of 6 cavities in the existing HINS beamline has been completed and beam measurements have been made. We present data showing the energy stability of the 7 mA proton beam accelerated through the six cavities from 2.5 MeV to 3.4 MeV..

## INTRODUCTION

As new and diverse applications for linear accelerators are developed, reducing construction and operating costs becomes a high design priority. RF power is one of the largest expenses for a linear accelerator, both for initial procurement and for operations. Most pulsed, linear accelerators utilize one klystron or IOT per accelerating structure. However, the cost per watt of pulsed power klystrons goes down for increasing power. A 5MW pulsed power klystron is much less expensive than 100 50kW klystrons. There can be substantial cost savings if the accelerator is powered by one, large RF source with a suitable RF power distribution system to multiple accelerating structures [1]. A cost effective means of manipulating the RF phase and amplitude for each structure at high power is needed to realize the potential benefit.

A proof-of-principle demonstration of this concept was carried out employing Ferrite Vector Modulators (FVM) [2] as the high power control device. FVMs are high power 90° hybrids with shorted, ferrite-loaded transmission lines on two ports. Phase and amplitude control is accomplished by adjusting the magnetic bias of each ferrite with a solenoidal coil.

A proton linac was constructed with a Radio Frequency Quadrupole (RFQ) and six copper cavities. All of the RF components were driven with a single 325 MHz, 2.5MW pulsed klystron. The RF input to each of the copper cavities is controlled by Ferrite Vector Modulators, while the RFQ is controlled directly through the Low Level RF (LLRF) control system. Two beam position monitors downstream of the last cavity provide beam time-of-flight and phase measurements that give beam energy. Our figures of merit for the study are the stability of the beam energy out of the linac and the stability of RF phase and amplitude in each of the cavities.

#steimel@fnal.gov

## BEAM LINE DESIGN

The HINS accelerator starts with a 50 kV duoplasmatron proton source followed by a 2-solenoid scheme, Low Energy Beam Transport (LEBT) line [3]. The protons are then accelerated by the HINS RFQ from 50 keV to 2.5 MeV. The RFQ is followed by the six-cavity beam line.

### *Six-Cavity Beam Line*

The beam line contains six room-temperature cavities: two pill-box style buncher cavities and four triple-spoke resonators. Buncher cavities are tuned to provide longitudinal beam focusing and do not accelerate. The triple-spoke resonators, shown in Fig. 1, increase the beam energy from 2.5 MeV to 3.4 MeV. These cavities offer higher shunt impedance and smaller power consumption for a specified cavity field [4]. Quad triplets provide transverse focusing. The line is equipped with button Beam Position Monitors (BPMs), which were primarily used to measure variations of beam energy, employing time-of-flight techniques.

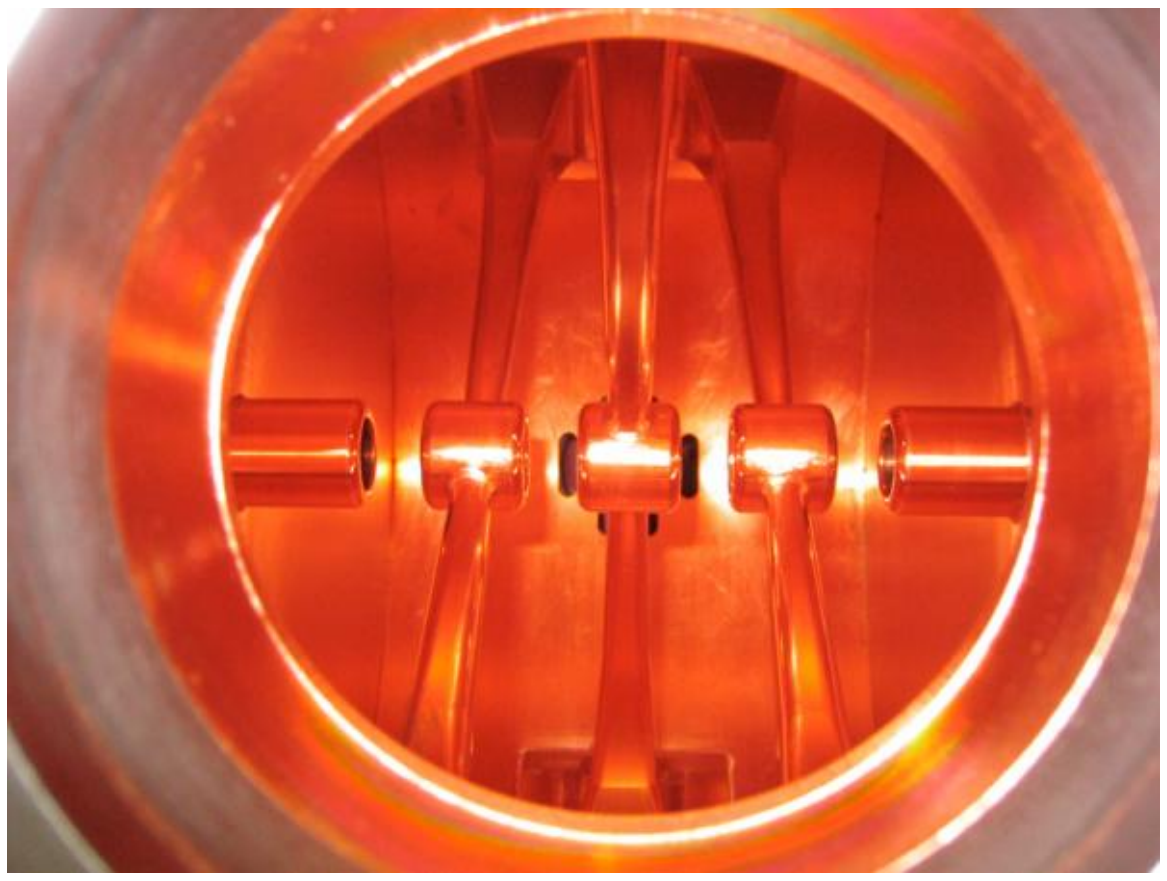

Figure 1: Inside of a copper, triple-spoke resonator [5].

The HINS diagnostic line follows the final buncher cavity. It consists of BPMs, a Fast Faraday Cup, a slit emittance scanner, three wire chambers, and a spectrometer magnet. The BPMs were used to verify the final beam energy by both time-of-flight techniques and measurements of the horizontal displacements downstream of the spectrometer magnet. Beam energy was also verified with wire scanners located at each end of the spectrometer magnet. Transverse beam properties were studied but were not the focus of this experiment.

### *RF Distribution*

The RF power distribution system evenly splits the klystron power between one branch feeding the RFQ and

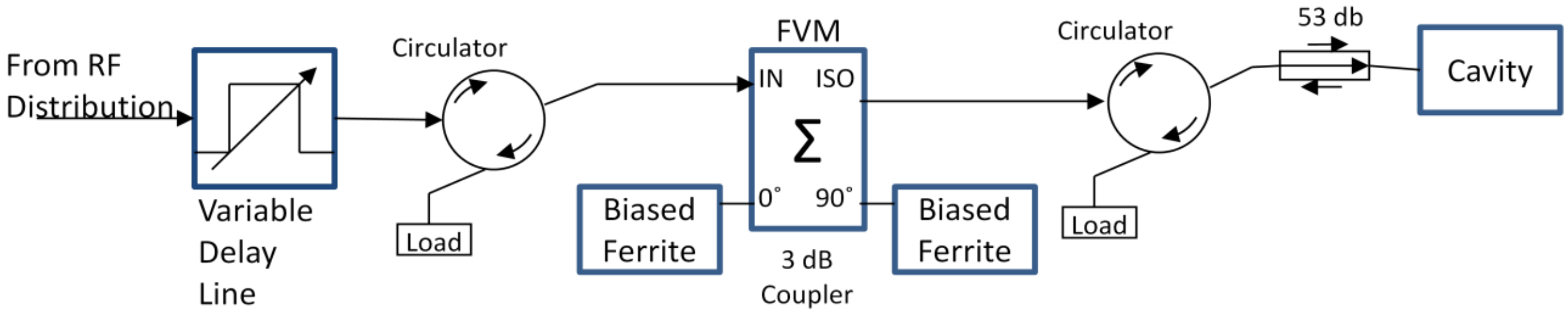

Figure 2: FVM circuit for each cavity.

a second branch feeding the other six cavities. The RFQ branch includes a mechanically variable, 0-3dB coupler that balances the beam line power requirement. The branch to the other six cavities includes a tree of 3dB, 4.77dB, and 6dB hybrids that bring the cavity power levels to within 1dB of their design power requirement

The RF input to each cavity is equipped with a FVM circuit shown in Fig. 2. Adjusting the bias current on the ferrites changes the delay of the shorted lines, changing the amplitude and/or phase of the output signal. FVMs have a limited range of operation and the amplitude and phase controls are coupled. For experimental flexibility, variable delay lines were installed to provide extended range of slow phase control and to maximize the useful FVM range.

## LLRF CONTROL SYSTEM

The HINS LLRF control system regulates the phase and amplitude of the RF field vectors of the RFQ and the six cavities. There is a traditional wide-band proportional and integral feedback control loop around the klystron and the RFQ. Because the RFQ is a low Q device, it behaves much like a resistive load. Therefore, by regulating the RFQ field, the klystron output to the six cavities is effectively regulated as well. This RFQ control loop greatly reduces errors from klystron modulator voltage variations and from changing beam current. There are several disturbances to the six cavities that must be corrected by the FVM controllers. These are static amplitude and phase errors, drifts in cavity resonance frequency, differences in cavity Qs, and variations in cavity beam loading. The FVMs are relatively slow devices (~27 kHz BW) so real-time feedback is difficult to stabilize. Instead, field errors in the cavities are corrected by an adaptive feed-forward system. This system has two parallel loops, one operating on the average (DC) error, and the other working on the AC component of the error using a time reversal filter algorithm. A lookup table, converting the phase and amplitude request to FVM solenoid current waveforms follows the controller section.

The wideband control loop is processed at the full 56 MHz sample rate of A/D converters, while the FVM control loops decimate the data to a 100 kHz rate, where it is processed and the controller output is written to a VXI 16 channel arbitrary waveform generator module. Regulation waveforms are recalculated at the machine repetition rate.

## RESULTS

Proton beam was successfully accelerated through the HINS beam line using a single klystron power source and FVMs for RF control. Beam current out of the RFQ was stabilized to about 7mA during the studies with about 6mA reaching the final beam dump. Two different tests were done to verify beam energy stability- a “DC Loop” only test and a test with the “AC Loop” also included.

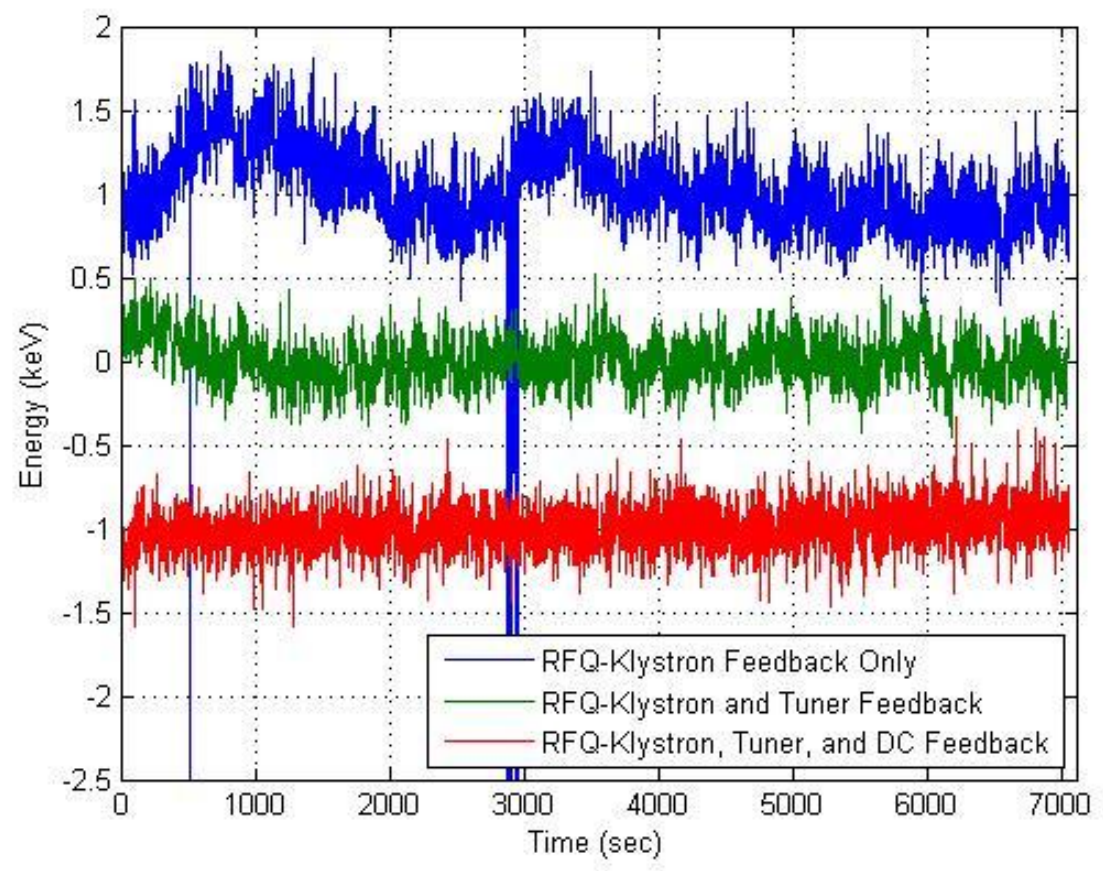

Figure 3: HINS proton beam energy variation over three, two hour time periods. The red and blue traces are offset by 1 keV for illustration. Each period uses different feedback schemes.

### *DC Control Loop Test*

The DC test verified the average energy stability of the pulse over a longer operating period. For this test, the beam pulse was 180µs long within a 200µs RF pulse. The beam repetition rate was 0.5 Hz. The first two BPMs immediately downstream of the last buncher cavity measured the 325 MHz component of the beam current. The phase of these signals are measured relative to the RFQ cavity probe and used to calculate changes in time-of-flight and beam energy. These phase detectors are sampled once per pulse, in the middle of the beam pulse.

Figure 3 shows the beam energy variation over three different, two-hour periods. In the first period, only the klystron-to-RFQ LLRF feedback system was active. In the second period, the individual cavity resonant frequency tuner feedback system was added. In the third period, the FVM DC loop was additionally activated. The plots are offset by 1 keV for easy comparison. The rms energy drift with all loops active is better than 1 keV and better than 1% of the available energy gain from the six cavities. The actual drift is likely smaller, obscured by the noise of the beam energy measurement.

### *DC and AC Control Loop Test*

The AC test, with DC loops activated, verified the stability of the cavity fields across longer pulses in the presence of beam loading. For this test, the pulse length was extended to 300μs with a 400μs RF pulse. Figures 4 and 5 show the magnitude and phase of spoke cavity #4 field probe signal as sampled by the LLRF system.

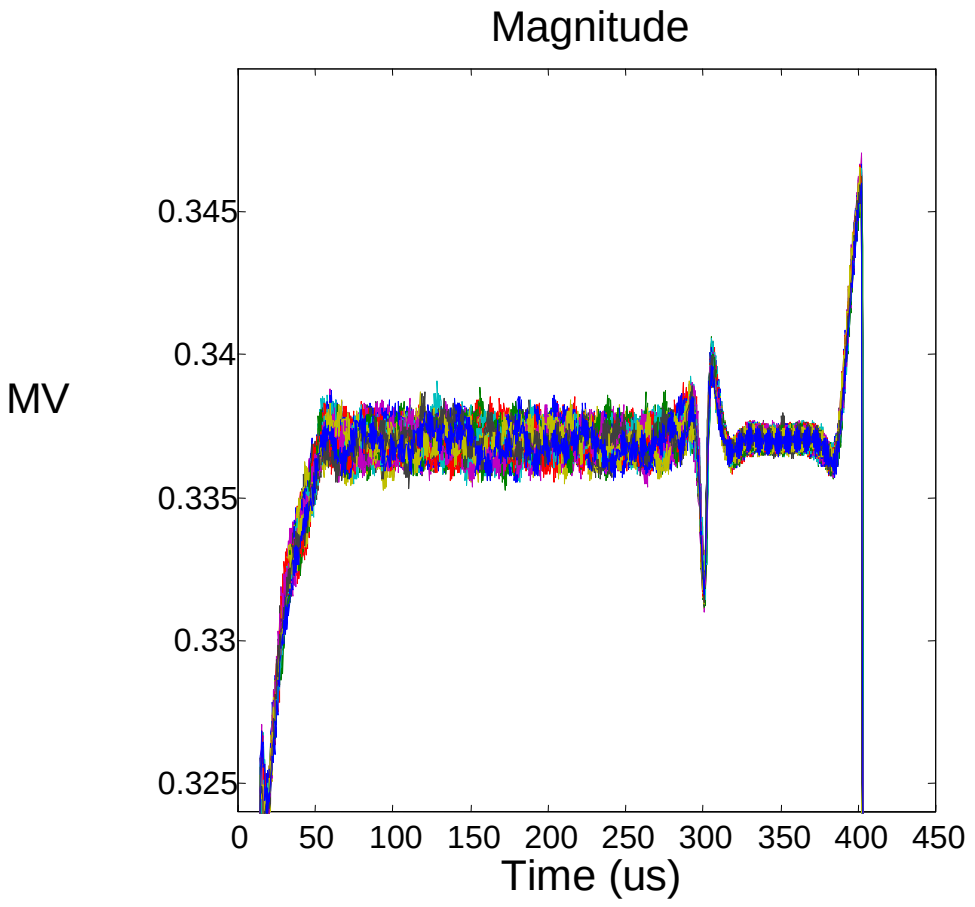

Figure 4: Plot showing 50 overlaid traces of spoke resonator cavity 4 field probe magnitude while accelerating beam.

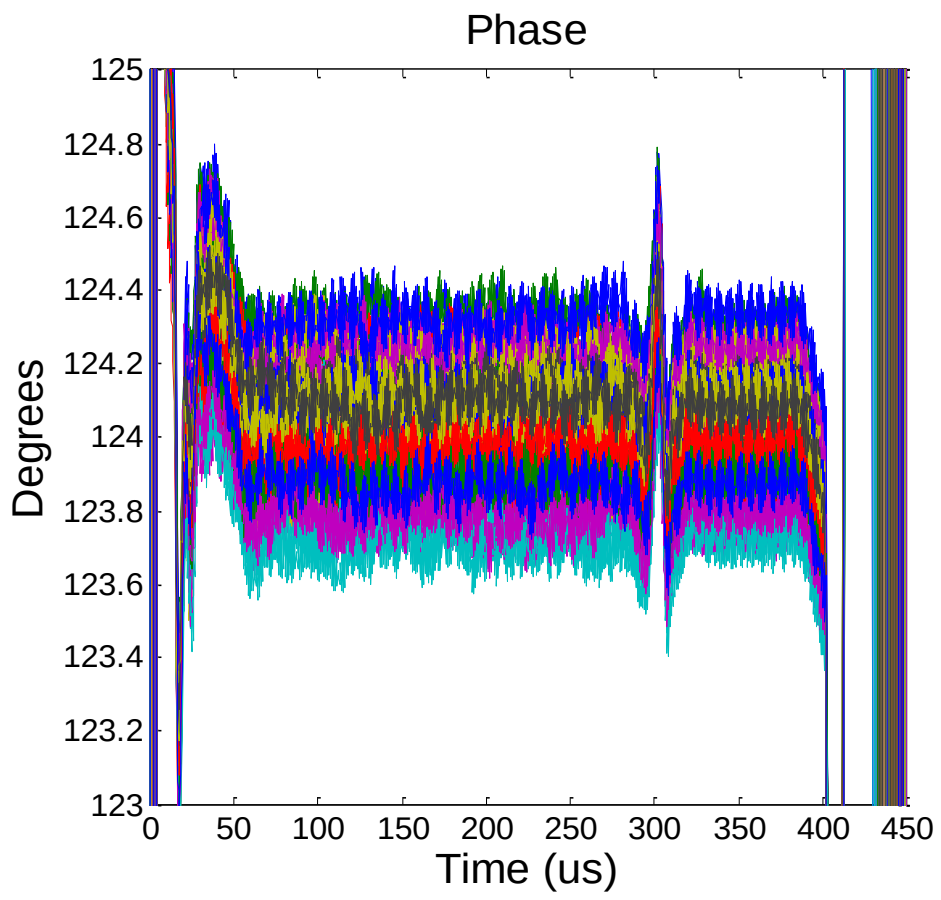

Figure 5: Plot showing 50 overlaid traces of spoke resonator cavity 4 field probe phase while accelerating beam.

Each plot shows 50 overlaid traces of consecutive beam pulses. The spike at 320μs shows where beam is stopped while the RF stays on. This gives a measure of the time response of the system. The time response was limited by the bandwidth of the FVMs and the 100 kHz DAC that drives the FVM current reference. Table 1 summarizes the cavity field regulation for all cavities, with and without AC feedback active. Cavity 1 was limited in its ability to regulate beam loading effects because it had the least available power overhead of any cavity and it ran out of dynamic range.

## CONCLUSIONS

This demonstration shows that FVMs can be used to maintain stable beam energy in a pulsed, proton accelerator. The tests show better than 1% amplitude regulation and 1degree phase regulation, both intra-pulse and pulse-to-pulse. Further study is needed to determine the cost effectiveness of a larger FVM controlled accelerator. Some extra components, not operationally required, were included in our RF distribution system for ease of commissioning and experimenting. This demonstration is also a first successful test of beam acceleration with spoke-style resonators.

Table 1: RMS Cavity Field Magnitude and Phase Variation Across the Pulse with Beam and RFQ-Klystron Feedback.

| **Description** | **FVM Control OFF** | | **FVM Control ON** | |
|---|---|---|---|---|
| | **Mag. (%)** | **Phase (deg)** | **Mag. (%)** | **Phase (deg)** |
| RFQ | 0.021 | 0.015 | 0.021 | 0.015 |
| Buncher 1 | 0.605 | 0.945 | 0.142 | 0.089 |
| Cavity 1* | 2.254 | 0.435 | 1.664 | 0.647 |
| Cavity 2 | 1.737 | 1.200 | 0.203 | 0.209 |
| Cavity 3 | 1.070 | 1.434 | 0.201 | 0.145 |
| Cavity 4 | 0.543 | 1.887 | 0.159 | 0.149 |
| Buncher 2 | 0.457 | 2.314 | 0.190 | 0.113 |

*FVM control dynamic range limit reached.